\documentclass[twocolumn]{el-author}

\newcommand{\ua}{\uparrow}
\newcommand{\nc}{\newcommand}
\nc{\da}{\downarrow} \nc{\hc}{\hat{c}} \nc{\hS}{\hat{S}}
\nc{\bra}{\langle} \nc{\ket}{\rangle} \nc{\eq}{equation (\ref}
\nc{\h}{\hat} \nc{\hT}{\h{T}}\nc{\be}{\begin{eqnarray}}
\nc{\ee}{\end{eqnarray}}\nc{\rd}{\textrm{d}}\nc{\e}{eqnarray}\nc{\hR}{\hat{R}}\nc{\Tr}{\mathrm{Tr}}
\nc{\tS}{\tilde{S}}\nc{\tr}{\mathrm{tr}}\nc{\8}{\infty}\nc{\lgs}{\bra\ua,\phi|}\nc{\rgs}{|\ua,\phi\ket}
\nc{\hU}{\hat{U}}\nc{\lfs}{\bra\phi|}\nc{\rfs}{|\phi\ket}\nc{\hZ}{\hat{Z}}\nc{\hd}{\hat{d}}\nc{\mD}{\mathcal{D}}
\nc{\bd}{\bar{d}}\nc{\bc}{\bar{c}}\nc{\mc}{\mathcal}\nc{\ea}{eqnarray}\nc{\mG}{\mathcal{G}}\nc{\bce}{\begin{center}}
\nc{\ece}{\end{center}}

\begin{document}

\title{Sharp inequality for $\ell_p$ quasi-norm and $\ell_q$-norm with $0<p\leq 1$ and $q>1$}

\author{Zenghui Zhang}

\abstract{
A sharp inequality for $\ell_p$ quasi-norm with $0<p\leq 1$ and $\ell_q$-norm with $q>1$ is derived, which shows that the difference between $\|\textbf{\textit{x}}\|_p$ and $\|\textbf{\textit{x}}\|_q$ of an $n$-dimensional signal $\textbf{\textit{x}}$ is upper bounded by the difference between the maximum and minimum absolute value in $\textbf{\textit{x}}$. The inequality could be used to develop new $\ell_p$-minimization algorithms.
}

\maketitle

\section{Introduction}
The problem of recovering a high-dimensional sparse signal from a few numbers of linear measurements has attracted much attention \cite{Donoho2006}\cite{Candes2005}. Let $\textbf{\textit{x}}=(x_1,x_2,...,x_n)\in\mathbb{R}^n$ be the signal we need to recover. We say $\textbf{\textit{x}}$ is $k$-sparse if it has no more than $k$ nonzero elements, i.e., $\|\textbf{\textit{x}}\|_0\leq k$. Let $\Phi\in\mathbb{R}^{m\times n}$ be the measurement matrix with $m<<n$. We have $\textbf{\textit{b}} = \Phi \textbf{\textit{x}}+\textbf{\textit{z}}$, where $\textbf{\textit{z}}\in\mathbb{R}^n$ is a vector of measurement errors and we assume that $\|\textbf{\textit{z}}\|_2\leq\varepsilon$. The sparse recovery problem is to reconstruct $\textbf{\textit{x}}$ based on $\textbf{\textit{b}}$ and $\Phi$. It can be solved by the following $\ell_0$-minimization
\begin{equation} \label{eq1}
(P_0)\ \ \min_\textbf{\textit{x}} \|\textbf{\textit{x}}\|_0,\ s.t.\ \|\textbf{\textit{b}}-\Phi \textbf{\textit{x}}\|_2\leq\varepsilon.
\end{equation}
However, $(P_0)$ is an NP-hard problem and therefore can not be solved efficiently \cite{Natarajan1995}. As alternative strategies, many substitution models for $(P_0)$ are proposed by replacing $\|\textbf{\textit{x}}\|_0$ with functions that evaluate the desirability of a would-be solution to $\textbf{\textit{b}} = \Phi \textbf{\textit{x}}$. Because of
\begin{equation}\label{eq2}
\|\textbf{\textit{x}}\|_0 = \lim_{p\rightarrow 0^{+}}\sum_{i=1}^n |x_i|^p = \lim_{p\rightarrow 0^{+}}\|\textbf{\textit{x}}\|_p^p,
\end{equation}
the following $\ell_p$-minimization with $0<p\leq1$ is often used \cite{Gribonval2003}\cite{Chartrand2007}\cite{Foucart2009}
\begin{equation} \label{eq3}
(P_p)\ \ \min_\textbf{\textit{x}} \|\textbf{\textit{x}}\|_p,\ s.t.\ \|\textbf{\textit{b}}-\Phi \textbf{\textit{x}}\|_2\leq\varepsilon.
\end{equation}

\begin{figure}[h]
\centerline{\includegraphics[width=3in]{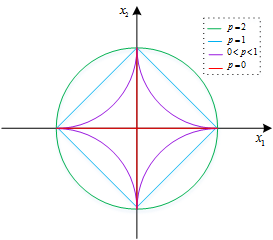}}\vspace{-0.3cm}
\caption{The behavior of $\ell_p$-norm for various values of $p$.}
\label{fig1}
\end{figure}

The behavior of different norms is illustrated in Fig. 1. Researchers show that the $\ell_p$-minimization with $0<p<1$ could recover a sparse signal with fewer measurements than the traditional used $\ell_1$-minimization. A central problem in $(P_p)$ is to find the relationship between $\|\textbf{\textit{x}}\|_p$ and $\|\textbf{\textit{x}}\|_2$. In 2010, Cai T., Wang L., and Xu G. \cite{Cai2010IT} gave a norm inequality for $\ell_1$ and $\ell_2$ as
\begin{equation} \label{eq4}
0\leq \|\textbf{\textit{x}}\|_2-\frac{\|\textbf{\textit{x}}\|_1}{\sqrt{n}} \leq
\frac{\sqrt{n}}{4}\Big(\max_{1\leq i\leq n}|x_i|-\min_{1\leq i\leq
n}|x_i|\Big).
\end{equation}

In this letter, a sharp inequality for $\ell_p$ and $\ell_q$ with $0<p\le1$ and $q>1$ is presented, which results in a new inequality for $\ell_p$ and $\ell_2$ as
\begin{equation} \label{eq5}
0\leq \|\textbf{\textit{x}}\|_2-n^{1/2-1/p}\|\textbf{\textit{x}}\|_p \leq
c_{p,2}
\sqrt{n}\Big(\max_{1\leq i\leq n}|x_i|-\min_{1\leq i\leq
n}|x_i|\Big),
\end{equation}
where
\begin{equation} \label{eq6}
c_p=\Big(1-\frac{p}{2}\Big)\Big(\frac{p}{2}\Big)^{\frac{p}{2-p}}.
\end{equation}

\section{Norm inequality for $\ell_p$ and $\ell_q$}

First, we give a lemma below, which will be used to prove the main result of this letter.

\textbf{Lemma 1}
Let
\[
s(x,y)=\Big(kx^q+(n-k)y^q\Big)^{1/q}-n^{1/q-1/p}\Big(kx^p+(n-k)y^p\Big)^{1/p},
\]
where $x>y\geq0$, $0<p\leq1$, $q>1$, $n, k$ are positive integers, and $1\leq k<n$. We have
\[
s(x,y)\leq s(x-y,0).
\]

\textbf{[Proof]} Let
\[
\begin{split}
h(t)&=n^{-1/q}s(x-t,y-t)\\
&=\Big(\frac{k}{n}(x-t)^q+\frac{n-k}{n}(y-t)^q\Big)^{1/q}\\
&-\Big(\frac{k}{n}(x-t)^p+\frac{n-k}{n}(y-t)^p\Big)^{1/p}
\end{split}
\]
with $0\leq t\leq y$. Its derivative about $t$ is
\[
\begin{split}
h'(t)&=-\Big[\frac{k}{n}\Big(\frac{x-t}{y-t}\Big)^q+\frac{n-k}{n}\Big]^{1/q-1}\Big[\frac{k}{n}\Big(\frac{x-t}{y-t}\Big)^{q-1}+\frac{n-k}{n}\Big]\\
     &+\Big[\frac{k}{n}\Big(\frac{x-t}{y-t}\Big)^p+\frac{n-k}{n}\Big]^{1/p-1}\Big[\frac{k}{n}\Big(\frac{x-t}{y-t}\Big)^{p-1}+\frac{n-k}{n}\Big].\\
\end{split}
\]
Consider the function
$g(x,q)=\big(ax^q+1-a\big)^{1/q-1}\big(ax^{q-1}+1-a\big)$ with
$x\geq1$ and $0\leq a<1$. We have
\[
g'(x,q)=(1-q)a(1-a)x^{q-2}(x-1)\big(ax^q+1-a\big)^{1/q-2}.
\]
For $q>1$, $g'(x,q)\leq0$ and $g(x,q)\leq g(1,q)=1$. For $0<p\leq1$, $g'(x,p)\geq0$ and $g(x,p)\geq g(1,p)=1$. Therefore, we
have
\[
h'(t) =
g\Big(\frac{x-t}{y-t},p\big)-g\Big(\frac{x-t}{y-t},q\Big)\geq0.
\]
Thus, $h(t)$ is increasing with $t$, which yields $s(x,y)\leq s(x-y,0)$. 
\newline

\textbf{Theorem 1} For any $\textbf{\textit{x}}=(x_1,x_2,...,x_n)\in\mathbb{R}^n$, $0<p\leq1$ and $q>1$, we have
\begin{equation} \label{eq8}
0\leq\|\textbf{\textit{x}}\|_q-n^{1/q-1/p}\|\textbf{\textit{x}}\|_p \leq
n^{1/q}c_{p,q}\Big(\max_{1\leq i\leq n}|x_i|-\min_{1\leq i\leq n}|x_i|\Big),
\end{equation}
with $c_{p,q}=(1-p/q)(p/q)^{p/(q-p)}$. The first equality holds if and only if $|x_1|=|x_2|=...=|x_n|$. The second equality holds if and only if $|x_1|=|x_2|=...=|x_n|$, or $m=n(p/q)^{pq/(q-p)}$ is a positive integer and $x$ satisfies $|x_{i_1}|=|x_{i_2}|=...=|x_{i_m}|$ for some $1\leq i_1<i_2<...<i_m\leq n$ and $x_k=0$ for $k\notin \{i_1,i_2,...,i_m\}$.

\textbf{[Proof]}
(1) The first part of the inequality.

Suppose $x_i\geq0,i=1,...,n$. We consider the function
\[
f(p) = \log{\Big(n^{-1/p}\|\textbf{\textit{x}}\|_p\Big)} =
\frac{1}{p}\log{\Big(\frac{1}{n}\sum_{i=1}^nx_i^p\Big)}
\]
Its derivative about $p$ is
\[
\begin{split}
f'(p) &=
-\frac{1}{p^2}\log{\Big(\frac{1}{n}\sum_{i=1}^nx_i^p\Big)}+\frac{1}{p}\frac{\frac{1}{n}\sum_{i=1}^n x_i^p\log{x_i}}{\frac{1}{n}\sum_{i=1}^nx_i^p}\\
      &=-\frac{1}{p^2\frac{1}{n}\sum_{i=1}^nx_i^p}\bigg[\Big(\frac{1}{n}\sum_{i=1}^nx_i^p\Big)\log{\Big(\frac{1}{n}\sum_{i=1}^nx_i^p\Big)}-\frac{1}{n}\sum_{i=1}^n
      x_i^p\log{x_i^p}\bigg]
\end{split}
\]
Let $g(x)=x\log{x}, x>0$. We have $g''(x)=1/x>0$, which means that $g(x)$ is strictly concave. Thus,
\[
\begin{split}
\Big(\frac{1}{n}\sum_{i=1}^nx_i^p\Big)\log{\Big(\frac{1}{n}\sum_{i=1}^nx_i^p\Big)} &= g\Big(\frac{1}{n}\sum_{i=1}^nx_i^p\Big) \\
\leq \frac{1}{n}\sum_{i=1}^ng(x_i^p) &=\frac{1}{n}\sum_{i=1}^n x_i^p\log{x_i^p}
\end{split}
\]
Therefore, $f'(p)\geq0$ and $f(p)$ is increasing with $p$. If
$0<p<q$, we have
\[
\|\textbf{\textit{x}}\|_q-n^{1/q-1/p}\|\textbf{\textit{x}}\|_p =n^{1/q}(f(q)-f(p))\geq0
\]
The equality is attained if and only if $x_1=x_2=...=x_n$.

(2) The second part of the inequality.

It is obvious that the result holds if $|x_1|=|x_2|=...=|x_n|$.
Without loss of generality, we assume that $x_1\geq x_2\geq ...\geq
x_n\geq0$ and not all $x_i$ are equal. Let
\[
f(\textbf{\textit{x}})=\|\textbf{\textit{x}}\|_q-n^{1/q-1/p}\|\textbf{\textit{x}}\|_p.
\]
We have
\[
\frac{\partial f}{\partial
x_i}=x_i^{q-1}\|\textbf{\textit{x}}\|_q^{1-q}-n^{1/q-1/p}x_i^{p-1}\|\textbf{\textit{x}}\|_p^{1-p}
\]
and
\[
\begin{split}
\frac{\partial^2 f}{\partial
x_i^2}&=(q-1)x_i^{q-2}\Big(\sum_{i=1}^nx_i^q\Big)^{1/q-1}\Big(1-\frac{x_i^q}{\sum_{i=1}^nx_i^q}\Big)\\
&+n^{1/q-1/p}(1-p)x_i^{p-2}\Big(\sum_{i=1}^nx_i^p\Big)^{1/p-1}\Big(1-\frac{x_i^p}{\sum_{i=1}^nx_i^p}\Big).
\end{split}
\]
If $0<p\leq1, q>1$, $\frac{\partial^2 f}{\partial x_i^2}\geq0$ which
shows that $f(\textbf{\textit{x}})$ is convex. Therefore, if we fix $x_1$ and $x_n$,
$f(\textbf{\textit{x}})$ must achieve its maximum on the borders. This implies that
the maximum has the form of $x_1=x_2=...=x_k$ and
$x_{k+1}=x_{k+2}=...=x_n$ for some $1\leq k<n$. Thus
\[
f(\textbf{\textit{x}})\leq\Big(kx_1^q+(n-k)x_n^q\Big)^{1/q}-n^{1/q-1/p}\Big(kx_1^p+(n-k)x_n^p\Big)^{1/p}.
\]
By Lemma 1, we have
\[
f(\textbf{\textit{x}})\leq k^{1/q}(x_1-x_n)-n^{1/q-1/p}k^{1/p}(x_1-x_n).
\]
Treat the right-hand side of the above as a function of $k$ for $k\in(0,n)$
\[
l(k)=k^{1/q}(x_1-x_n)-n^{1/q-1/p}k^{1/p}(x_1-x_n).
\]
By taking the derivative, we have $l'(k)=0$ if
$k=n\big(p/q\big)^{pq/(q-p)}$. Therefore
\[
f(\textbf{\textit{x}})\leq l(k)\leq
n^{1/q}\Big(1-\frac{p}{q}\Big)\Big(\frac{p}{q}\Big)^{\frac{p}{q-p}}(x_1-x_n).
\]
Proof of Theorem 1 is completed. 
\newline

\section{Discussions}
Consider the inequality of (\ref{eq8}), if we define the normalized $\ell_p$ quasi-norm of $\textbf{\textit{x}}$ as
\[
\|\textbf{\textit{x}}\|_{\bar{p}} = \Big(\frac{|x_1|^p+|x_2|^p+...+|x_n|^p}{n}\Big)^{\frac{1}{p}},
\]
we have
\[
0\leq \|\textbf{\textit{x}}\|_{\bar{q}}-\|\textbf{\textit{x}}\|_{\bar{p}}\leq c_{p,q}\Big(\max_{1\leq i\leq n}|x_i|-\min_{1\leq i\leq n}|x_i|\Big).
\]
Thus, the constant $c_{p,q}$ is critical for measuring the sharpness of the inequality. The changing of $c_{p,q}$ with $0<p\leq 1$ for various values of $q$ is illustrated in Fig. 2.

\begin{figure}[h]
\centerline{\includegraphics[width=3in]{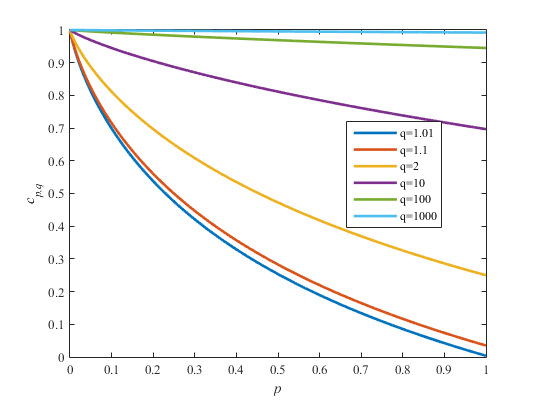}}\vspace{-0.3cm}
\caption{The changing of $c_{p,q}$ with $p$ for various values of $q$.}
\label{fig2}
\end{figure}

From Fig. 2, we can get the following results for $c_{p,q}$.

(1) $0\leq c_{p,q} \leq 1$, which means that the difference between $\|\textbf{\textit{x}}\|_{\bar{q}}$ and $\|\textbf{\textit{x}}\|_{\bar{p}}$ is no more than the difference between the maximum and minimum absolute value in $\textbf{\textit{x}}$. Also, we have $\lim_{p\rightarrow 0}c_{p,q}=0$ and $\lim_{q\rightarrow +\infty}c_{p,q}=1$. Therefore, the inequality of (\ref{eq8}) is very sharp.

(2) For every fixed $q$, $c_{p,q}$ is monotonously decreasing with $p$. It is easy to prove because that $1-p/q$ and $(p/q)^{p/(q-p)}$ are all monotonously decreasing with $p$. If we consider the function of $l(p) = p/(q-p)ln(p/q)$ , we have $l'(p)=q/(q-p)^2\big(1+ln(p/q)-p/q\big)<0$ for $0<p\leq 1$ and $q>1$.

(3) For every fixed $p$, $c_{p,q}$ is monotonously increasing with $q$. If we consider the function of $l(q)=p/(q-p)ln(p/q)$, we have $l'(q)=p/(q-p)^2\big(-1-ln(p/q)+p/q\big)>0$ for $0<p\leq 1$ and $q>1$.

A direct consequence of Theorem 1 is that for any $\textbf{\textit{x}}\in\mathbb{R}^n$ and $0<p\leq1$,
\[
0\leq\|\textbf{\textit{x}}\|_2-n^{1/2-1/p}\|\textbf{\textit{x}}\|_p \leq
c_{p,2}\sqrt{n}\Big(\max_{1\leq
i\leq n}|x_i|-\min_{1\leq i\leq n}|x_i|\Big),
\]
where $c_{p,2}$ is defined in (\ref{eq6}).

\section{Conclusion}
A new inequality for $\ell_p$-norm and $\ell_q$ quasi-norm of an $n$-dimensional signal is proposed, and the conditions that the inequality holds are given in the case where $0<p\leq 1$ and $q>1$. Analysis shows that the new inequality is very sharp. Because the relationship between $\ell_p$ quasi-norm and $\ell_2$-norm is critical for the research of $\ell_p$-minimization problems, the new inequality could be used to develop new $\ell_p$-minimization algorithms. The generalization of the norm inequality for arbitrary $0<p<q$ will be studied in the future.

\vskip5pt

\noindent Zenghui Zhang(\textit{School of Electronics, Information, and Electrical Engineering, Shanghai Jiao Tong University, P.R.China})
\vskip3pt

\noindent E-mail: zenghui.zhang@sjtu.edu.cn

\end{document}